\documentclass[aps,prl,amsmath,amssymb,reprint]{revtex4-1}
\usepackage[utf8]{inputenc}
\usepackage[T1]{fontenc}
\usepackage{lmodern}
\usepackage{graphicx}
\usepackage{hyperref}

\begin{document}

\title{Driven-state relaxation of a coupled qubit-defect system in spin-locking measurements} 

\author{Leonid V. Abdurakhimov}
\email{leonid.abdurakhimov.nz@hco.ntt.co.jp}
\affiliation{NTT Basic Research Laboratories, NTT Corporation, 3-1 Morinosato-Wakamiya, Atsugi, Kanagawa 243-0198, Japan}

\author{Imran Mahboob}
\affiliation{NTT Basic Research Laboratories, NTT Corporation, 3-1 Morinosato-Wakamiya, Atsugi, Kanagawa 243-0198, Japan}

\author{Hiraku Toida}
\affiliation{NTT Basic Research Laboratories, NTT Corporation, 3-1 Morinosato-Wakamiya, Atsugi, Kanagawa 243-0198, Japan}

\author{Kosuke Kakuyanagi}
\affiliation{NTT Basic Research Laboratories, NTT Corporation, 3-1 Morinosato-Wakamiya, Atsugi, Kanagawa 243-0198, Japan}

\author{Yuichiro Matsuzaki}
\altaffiliation[Present address: ]{Nanoelectronics Research Institute, National Institute of Advanced Industrial Science and Technology (AIST), 1-1-1 Umezono, Tsukuba, Ibaraki 305-8568, Japan}

\author{Shiro Saito}
\affiliation{NTT Basic Research Laboratories, NTT Corporation, 3-1 Morinosato-Wakamiya, Atsugi, Kanagawa 243-0198, Japan}

\date{\today}

\begin{abstract}
It is widely known that spin-locking noise-spectroscopy is a powerful technique for the characterization of \textit{low-frequency} noise mechanisms in superconducting qubits. Here we show that the relaxation rate of the driven spin-locking state of a qubit can be significantly affected by the presence of an off-resonant \textit{high-frequency} two-level-system defect. Thus, both low- and high-frequency defects should be taken into account in the interpretation of spin-locking measurements and other types of driven-state noise-spectroscopy. 
\end{abstract}

\maketitle 

The field of gate-based quantum computing using superconducting qubits is rapidly developing\,\cite{Arute2019}. However, the realization of a fault-tolerant quantum computer remains a challenging task: the implementation of a reasonably robust logical qubit would require an overhead of order $10^3$ to $10^4$ physical qubits with per-gate error rates $p\approx10^{-3}$\,\cite{Fowler2012}. Lowering per-operation error rates would significantly reduce the overhead requirements, and, therefore, it is important to understand noise-induced error mechanisms during free and driven evolution of qubits. Free-evolution qubit relaxation has been extensively studied, and standard protocols for measurements of energy relaxation and dephasing rates --- including spin-echo, Ramsey, and dynamical decoupling methods --- have been widely used\,\cite{Ithier2005,Bylander2011,Yan2016}. As for the decoherence of a superconducting qubit during driven evolution, qubit relaxation mechanisms in Rabi and spin-locking measurements were initially analyzed in Refs.\,\cite{Smirnov2003,Ithier2005}. The Rabi spectroscopy was later used in experimental studies of noise characteristics in superconducting flux qubits\,\cite{Bylander2011,Gustavsson2012,Yoshihara2014}, and in a qubit-fluctuator system\,\cite{Lisenfeld2010}. More recently, spin-locking spectroscopy has been demonstrated to be a powerful tool for noise characterization in superconducting qubits\,\cite{Yan2013}, which can be used to detect low-frequency defects\,\cite{Yan2013}, distinguish between coherent and thermal photon noise~\cite{Yan2018}, and measure low-frequency noises in multi-qubit systems~\cite{Luepke2019,Sung2020}.

Spin-locking noise spectroscopy of a superconducting qubit is based on the measurements of the qubit evolution driven by a spin-locking pulse sequence originally developed in NMR studies\,\cite{Yan2013}. In the Bloch-sphere representation in a rotating frame, a qubit state is associated with a fictitious spin-$\frac{1}{2}$ state, and a spin-locking measurement is described by the sequence of three consecutive pulses $(-\pi/2)_Y - \text{SL}_X - (-\pi/2)_Y $. The first pulse rotates the spin around the $y$-axis by a $-\pi/2$ angle resulting in the spin oriented along the $x$-axis. The second pulse --- a so-called spin-locking pulse --- is a long pulse with the variable amplitude and duration which is applied along the $x$-axis. Driven by the spin-locking pulse, the spin precesses around the $x$-axis at the Rabi frequency $\Omega_\text{R}$ (determined by the pulse amplitude), and, thus, the spin is effectively ``locked'' along the $x$-axis. The third pulse aligns the spin along the $z$-axis which allows one to measure the qubit state (e.g., by dispersive readout). In spin-locking measurements, both the amplitude and duration $\tau$ of the spin-locking pulse are varied, and, at a given amplitude,  the relaxation rate is extracted from an exponential fit of the qubit excited-state population decay, $P_\text{e} =(1 + \exp(-\Gamma_{1\rho}\tau))/2$. According to the model of generalized Bloch equations (GBE)~\cite{Smirnov2003,Ithier2005,Yan2013}, the relaxation rate $\Gamma_{1\rho}$ and relaxation time $T_{1\rho}$ of the qubit driven-state in the rotating frame (hence, the subscript symbol ``$\rho$'' is used conventionally) are given by:
\begin{equation}
\Gamma_{1\rho} = \frac{1}{T_{1\rho}} = \frac{1}{2} \Gamma^\prime_1 + \Gamma_{\Omega}(\Omega_\text{R}),
\label{eq:eq1}
\end{equation}
where the relaxation rate $\Gamma_{\Omega}(\Omega_\text{R})$ is related to the low-frequency longitudinal noise at the Rabi frequency $\Omega_\text{R}$, while the relaxation rate $\Gamma^\prime_1$ is usually assumed to be determined by the high-frequency transverse noise at the qubit frequency $\omega_{\text{q}}$ ($\omega_{\text{q}} \gg \Omega_\text{R}$)~\cite{Ithier2005,Yan2013, Yan2018}:
\begin{equation}
\Gamma^\prime_1 \approx \Gamma_1 (\omega_{\text{q}})=(T_1(\omega_{\text{q}}))^{-1},
\label{eq:approximation}
\end{equation}
where $T_1$ is the qubit energy-relaxation time. However, it is known that the exact equation for $\Gamma^\prime_1$ is given by\,\cite{Geva1995,Smirnov2003,Yan2013,first_note}:
\begin{equation}
\Gamma^\prime_1 = \frac{1}{2}(\Gamma_1 (\omega_{\text{q}}-\Omega_\text{R})+\Gamma_1 (\omega_{\text{q}}+\Omega_\text{R})),
\label{eq:exact}
\end{equation}
and, therefore, Eq.~(\ref{eq:approximation}) is valid only if $\Gamma_1 (\omega_{\text{q}}) \approx \Gamma_1 (\omega_{\text{q}} \pm \Omega_\text{R})$. Equation~(\ref{eq:exact}) is usually disregarded, and it is assumed that low-frequency noise sources can be unambiguously characterized from $T_{1\rho}$ and $T_1$ measurements at the qubit frequency $\omega_{\text{q}}$ by extracting the relaxation rate at the Rabi frequency through the relation $\Gamma_{\Omega}(\Omega_\text{R}) =  \Gamma_{1\rho}- \Gamma_1 (\omega_{\text{q}})/2$. For example, spectral features observed in the MHz frequency range in spin-locking experiments with a superconducting qubit were attributed to low-frequency fluctuators\,\cite{Yan2013}.
 
In this Rapid Communication, we present results of spin-locking measurements of a superconducting flux qubit coupled to off-resonant \textit{high-frequency} two-level-system (TLS) defects. We observed spectral features in the driven-state relaxation time $T_{1\rho}$ which were caused by the interaction between the qubit with the frequency $\omega_{\text{q}}$ and defects with the frequencies $\omega_{\text{TLS}}$ matching the conditions $\omega_{\text{TLS}}=\omega_{\text{q}} - \Omega_\text{R}$ or $\omega_{\text{TLS}}=\omega_{\text{q}} + \Omega_\text{R}$. Thus, both low-frequency and high-frequency noise sources should be taken into account when interpreting results of spin-locking noise-spectroscopy measurements.

\begin{figure}[t!]
    \centering
    \includegraphics{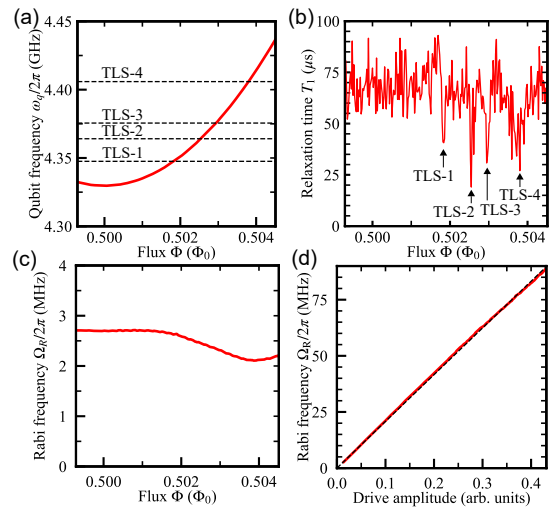}
    \caption{(color online) (a) The qubit frequency as a function of the applied magnetic flux. Dashed lines correspond to positions of TLS defects determined from $T_1$ measurements. (b) The energy-relaxation time $T_1$ of the qubit. Markers show approximate positions of four TLS defects resonantly coupled to the qubit. (c) The frequency of Rabi oscillations at a given drive amplitude. (d) The frequency of Rabi oscillations at the optimal bias point $\Phi = 0.5\,\Phi_0$ as a function of the drive amplitude. The black dashed line corresponds to a linear fit.}
    \label{fig1}
\end{figure}

Our sample is a capacitively-shunted (c-shunt) superconducting flux qubit embedded in a 3D microwave cavity. Detailed information about the qubit design and experimental setup can be found in Refs.~\cite{Abdurakhimov2019,SM}. In the reported spin-locking experiments, a commercial high-power microwave amplifier from Spacek Labs was used to drive the qubit\,\cite{SM}. The qubit was measured by dispersive readout. The cavity frequency was $\omega_\text{c}/2\pi \approx 8.2185$\,GHz, the cavity linewidth was $\kappa/2\pi \approx 1$\,MHz, and the dispersive frequency pull was $2\chi/2\pi \approx 1.4$\,MHz. The qubit frequency measured as a function of the applied magnetic flux $\Phi$ is shown in Fig.\,\ref{fig1}(a). The qubit frequency was $\omega_{\text{q}}/2\pi\approx 4.330$\,GHz at the optimal flux bias point $\Phi=0.5\,\Phi_0$, where $\Phi_0$ is the magnetic flux quantum. In a separate two-tone measurement, the qubit anharmonicity was found to be about 0.8\,GHz.  At the optimal bias point, qubit decoherence times $T_{2\text{E}}\approx 6$\,$\mu$s and $T_{2\text{R}}\approx 5$\,$\mu$s were measured using spin-echo and Ramsey pulse sequences, respectively. In comparison with previous measurements of the sample\,\cite{Abdurakhimov2019}, the qubit frequency was slightly lower due to the thermal cycling between the experimental runs, and the reduction of decoherence times $T_{2\text{E}}$ and $T_{2\text{R}}$ was caused by the amplitude and phase noises induced by the high-power amplifier\,\cite{SM}. 

The magnetic-flux dependence of the qubit energy-relaxation time $T_1$ is shown in Fig.\,\ref{fig1}(b). Variations of the $T_1$ values in the range 50--90 $\mu$s were related to the temporal variations reported previously\,\cite{Abdurakhimov2019} which were caused by quasiparticle tunneling\,\cite{Yan2016,Vepsaelaeinen2020,Cardani2020} and a bath of background TLS defects\,\cite{Mueller2015,Klimov2018,Burnett2019}. In addition, we observed four pronounced dips which can be attributed to a resonant coupling between the qubit and a subset of distinct TLS defects denoted as TLS-1, TLS-2, TLS-3, and TLS-4. The dependence of the energy-relaxation rate on parameters of a coupled qubit-defect system is non-trivial\,\cite{Paladino2010,Paladino2010a,Gustavsson2012a}. Qualitatively, the presence of observed TLS defects did not cause avoided crossings in the qubit spectrum, and, hence, the qubit-defect couplings were weak which was in contrast with the previous works on strongly-coupled qubit-defect systems\,\cite{Simmonds2004,Martinis2005,Lupascu2009,Lisenfeld2010,Gustavsson2012a,Lisenfeld2015,Bilmes2020}. 

Figure\,\ref{fig1}(c) shows the frequency  of Rabi oscillations at a given drive amplitude as a function of the applied magnetic flux. The decrease of the Rabi frequency at $\Phi\approx 0.504 \,\Phi_0$ was probably due to the coupling to the bath of high-frequency TLS defects. The dependence of the Rabi frequency on the drive amplitude at the optimal bias point $\Phi=0.5\,\Phi_0$ is shown in Fig.\,\ref{fig1}(d).

\begin{figure}[t!]
    \centering
    \includegraphics{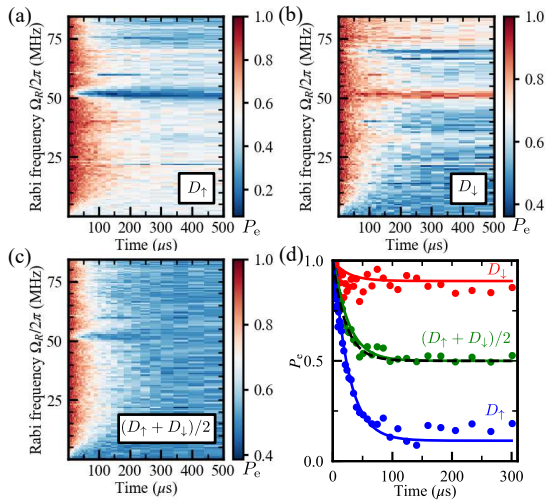}
    \caption{(color online) Results of spin-locking spectroscopy at the optimal point $\Phi=0.5\,\Phi_0$. (a) The data set $D_\uparrow$ represents results obtained using the pulse sequence $(-\frac{\pi}{2})_Y - \text{SL}_X - (-\frac{\pi}{2})_Y$. (b) The data set $D_\downarrow$ corresponds to results obtained using the pulse sequence $(+\frac{\pi}{2})_Y - \text{SL}_X - (+\frac{\pi}{2})_Y$. (c) The arithmetic mean $P=(D_\uparrow+D_\downarrow)/2$. (d) Driven-state relaxation of a coupled qubit-defect system at the Rabi frequency of 51.3\,MHz. Dots correspond to the traces taken from data sets shown in Figs.\,\ref{fig2}(a)--(c) at $\Omega_\text{R}/2\pi\approx 51.3$\,MHz. The black dashed line corresponds to the exponential fit of the data $P$ with the characteristic time $T_{1\rho} = 22$\,$\mu$s. Solid lines represent results of numerical simulations described in the text.}
    \label{fig2}
\end{figure}

Results of spin-locking measurements at the optimal bias point $\Phi=0.5\,\Phi_0$ are shown in Fig.\,\ref{fig2}. We used two types of spin-locking pulse sequences. The first one (labeled 'S1') was the standard sequence $(-\pi/2)_Y - \text{SL}_X - (-\pi/2)_Y $, in which the qubit was in the state $|0\rangle_x = (|0\rangle + |1\rangle)/\sqrt{2}$ (parallel to the $x$-axis) at the end of the first pulse. Here, the states $|0\rangle$ and $|1\rangle$ are eigenstates of the qubit Pauli operator $\sigma_z$ with eigenvalues $+1$ and $-1$ respectively. In experiments, we measured the probability to find the qubit in the excited state $|0\rangle$ in the end of the spin-locking sequence. Figure~\ref{fig2}(a) shows the data set $D_\uparrow$ obtained using the S1 sequence. The second sequence (labeled 'S2') was the modified spin-locking sequence $(+\pi/2)_Y - \text{SL}_X  - (+\pi/2)_Y $, in which the qubit was in the state $|1\rangle_x = (|0\rangle - |1\rangle)/\sqrt{2}$ (anti-parallel to the $x$-axis) at the end of the first pulse. The data set $D_\downarrow$ obtained using the S2 sequence is shown in Fig.\,\ref{fig2}(b). For the both types of pulse sequences, rectangular pulses were used. The $\pi/2$-pulse duration was chosen to be long ($\approx$\,100\,ns) to minimize spectral widths of the $\pi/2$-pulses. The amplitude and duration of $\text{SL}_X$-pulses were varied. Figure~\ref{fig2}(c) shows the arithmetic mean $P=(D_\uparrow+D_\downarrow)/2$ which was used for estimations of the relaxation time~$T_{1\rho}$. 

\begin{figure}[t!]
    \centering
    \includegraphics{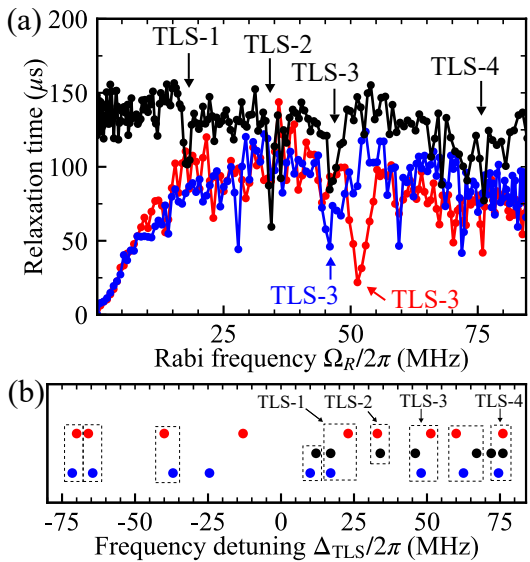}
    \caption{(color online) (a) The relaxation time $T_{1\rho}$ of the driven qubit state. Red dots represent $T_{1\rho}$ at the optimal bias point $\delta \Phi =0$ (the qubit frequency was $\omega_{\text{q}}/2\pi \approx 4.330$\,GHz). Blue dots correspond to $T_{1\rho}$ at the flux bias detuned from the optimal point by $\delta \Phi = 7.5\times10^{-4}\,\Phi_0$ (the qubit frequency detuning was $\delta \omega_{\text{q}}/2\pi \approx 3$\,MHz). Black dots represent the estimated values of $(\Gamma^\prime_1/2)^{-1}$ at the optimal bias point. Markers show spectral features caused by the coupling between the qubit and TLS defects in corresponding measurements. (b) Frequencies of high-frequency TLS defects determined from $T_1$ (black dots) and $T_{1\rho}$ (red and blue dots) measurements. The horizontal axis represents the frequency detuning $\Delta_{\text{TLS}} = \omega_{\text{TLS}} - \omega_{\text{q}}$. Red dots correspond to spectral features inferred from the $T_{1\rho}$ raw data obtained at the the optimal point~[Fig.\,\ref{fig2}(a),(b)]. Blue dots represent defect positions determined from the $T_{1\rho}$ raw data obtained at the detuned bias\,\cite{SM}, with the corresponding values $\Delta_{\text{TLS}}$ being increased by the value $\delta \omega_{\text{q}}$. Dashed-line rectangles indicate spectral features caused by the same TLS defect.}
    \label{fig3}
\end{figure}

A distinct spectral feature was observed in spin-locking measurements at the Rabi frequency $\Omega_\text{R}/2\pi\approx 51.3$\,MHz~[Figs.\,\ref{fig2}(a)--(c)]. We suppose that it was caused by the coupling between the qubit and the TLS-3 defect. Following Refs.~\cite{Zagoskin2006,Lupascu2009,Gustavsson2012a,Yan2013}, we model the qubit-TLS system subjected to a microwave excitation by the  Hamiltonian $ H = H_\text{q} + H_\text{TLS} + H_\text{I} + H_\text{MW}$, where  $H_\text{q}  = (\hbar\omega_{\text{q}}/2) \sigma_z^{\text{(1)}}$ is the qubit Hamiltonian, $H_\text{TLS}  =  (\hbar\omega_{\text{TLS}}/2)  \sigma_z^{\text{(2)}}$ is the defect Hamiltonian, $H_\text{I}  = \hbar g \sigma_x^{\text{(1)}} \sigma_x^{\text{(2)}}$ is the interaction term, and  $H_\text{MW}  = \hbar \Omega_\text{R} \cos (\omega_{\text{MW}}t)\sigma_x^{\text{(1)}}$ is the microwave drive term. Here, $\sigma_{x,y,z}^{\text{(1)}}$ ($\sigma_{x,y,z}^{\text{(2)}}$) are Pauli operators for the qubit (the defect), $g$ is the coupling strength, and $\omega_{\text{MW}}$ corresponds to the frequency of the applied microwave drive.
In the case of the resonant driving ($\omega_{\text{q}} = \omega_\text{MW}$), the Hamiltonian can be written in the rotating wave approximation:
\begin{equation}
 H_{\text{R}}/\hbar = \frac{\Omega_\text{R} }{2} \sigma_x^\text{(1)} + \frac{\Delta_{\text{TLS}}}{2} \sigma_z^{\text{(2)}} + g  (\sigma_+^\text{(1)} \sigma_-^{\text{(2)}}  +  \sigma_-^\text{(1)} \sigma_+^{\text{(2)}}),
\end{equation}
where $\Delta_{\text{TLS}} = \omega_{\text{TLS}} - \omega_{\text{q}}$, and $\sigma_\pm^\text{(1,2)}=(\sigma_x^\text{(1,2)} \pm i \sigma_y^\text{(1,2)})/2$. Using QuTiP package\,\cite{Johansson2013}, we numerically simulated the dynamics of the system in the rotating frame by solving the Lindblad master equation for given energy-relaxation rates $\Gamma_{1}^\text{(1)}$ and $\Gamma_{1}^\text{(2)}$, and dephasing rates $\Gamma_{2}^\text{(1)}$ and $\Gamma_{2}^\text{(2)}$ of the qubit and the TLS defect, respectively (with the corresponding collapse operators $\sqrt{\Gamma_{1}^\text{(1)}}\,\sigma_-^\text{(1)}$, $\sqrt{\Gamma_{1}^\text{(2)}}\,\sigma_-^\text{(2)}$, $\sqrt{\Gamma_{2}^\text{(1)}/2}\,\sigma_z^\text{(1)}$, and $\sqrt{\Gamma_{2}^\text{(2)}/2}\,\sigma_z^\text{(2)}$).  Following Ref.\,\cite{Lisenfeld2016}, we used the values  $\Gamma_{1}^\text{(2)}= 10^6$\,s$^{-1}$ and $\Gamma_{2}^\text{(2)}=0$ for the defect relaxation rates, and, for simplicity, we assumed the qubit was free of pure dephasing, $\Gamma_{2}^\text{(1)}=0$. To simulate the S1 (S2) measurement, the defect was initialized in its ground state in the laboratory frame, and the expectation value $\langle\sigma_x^\text{(1)}\rangle$ as a function of the evolution time was calculated for the qubit initial state $|0\rangle_x$ ($|1\rangle_x$) and converted to the corresponding value of $D_\uparrow = (1+\langle\sigma_x^\text{(1)}\rangle)/2$ ($D_\downarrow = (1-\langle\sigma_x^\text{(1)}\rangle)/2$). We found that the observed spectral feature can be well reproduced using the following parameters: $\Omega_\text{R}/2\pi = \Delta_{\text{TLS}}/2\pi = 51.3$\,MHz, the coupling strength $g/2\pi = 28$\,kHz, and the qubit energy-relaxation rate $\Gamma_{1}^\text{(1)}= 1.5\times10^4$\,s$^{-1}$ [Fig.\,\ref{fig2}(d)]. Thus, conditions of the Purcell-like effect were fulfilled in the rotating frame: $\Gamma_{1}^\text{(1)} < g < \Gamma_{1}^\text{(2)}$. We assume that the qubit effective dissipation rate can be roughly estimated as $\Gamma_\text{P} \approx g^2/ \Gamma_{1}^\text{(2)} \approx 0.31\times10^5$\,s$^{-1}$\,\cite{Wood2014,Bienfait2016}, which is close to the relaxation rate value $T_{1\rho}^{-1}\approx 0.45 \times 10^5$\,s$^{-1}$ extracted from an exponential fit of the spin-locking data~$P$~[Fig.\,\ref{fig2}(d)]. Therefore, the phase cycling procedure described above --- averaging the data sets obtained using S1 and S2 sequences with alternating pulse phases ---  is a valid method to obtain the information on the driven-state relaxation rate. The $T_{1\rho}$ data, presented in Fig.\,\ref{fig3}(a), was extracted from the signal $P$ using exponential fitting. 

Besides the main feature at $\Omega_\text{R}/2\pi = 51.3$\,MHz, various additional spectral features were observed at other Rabi frequencies [Fig.\,\ref{fig2}(a)--(c) and Fig.\,\ref{fig3}(b)]. All features can be divided into two groups based on their ``polarity'' that is defined by the relation between the steady levels of the excited-state population $D_\uparrow$ and $D_\downarrow$ in S1 and S2 measurements, respectively\,\cite{SM}. For ``negative polarity'' features, such as the one observed at $\Omega_\text{R}/2\pi = 70$\,MHz in Figs.\,\ref{fig2}(a),(b), the condition $D_\uparrow > D_\downarrow$ was fulfilled. For those features, steady values of $\langle\sigma_x^\text{(1)}\rangle$ in the end of the spin-locking pulse were positive\,\cite{third_note}, which was caused by the heating by defects with $\Delta_{\text{TLS}} < 0$ according to our model\,\cite{second_note}. Other features, such as the main one observed at $\Omega_\text{R}/2\pi = 51.3$\,MHz, can be attributed to the ``positive polarity'' group for which the condition $D_\uparrow < D_\downarrow$ was realized. For those features, steady values of $\langle\sigma_x^\text{(1)}\rangle$ were negative, which was caused by the cooling of the qubit by defects with $\Delta_{\text{TLS}} > 0$. It should be mentioned that the pure qubit state $|0\rangle_x$ ($|1\rangle_x$) corresponded to the excited (ground) state in the rotating frame with $\langle\sigma_x^\text{(1)}\rangle=1$ ($\langle\sigma_x^\text{(1)}\rangle=-1$), and, therefore, the steady spin-locking state of the qubit without coupling to the TLS defect would be a completely mixed state of $|0\rangle_x$ and $|1\rangle_x$ states with $\langle\sigma_x^\text{(1)}\rangle=0$.

Figure~\ref{fig3}(a) shows relaxation times $T_{1\rho}$ as a function of the Rabi frequency at the optimal bias point and the flux bias detuned from the optimal point by $\delta \Phi = \Phi - 0.5\Phi_0 \approx 7.5 \times 10^{-4} \,\Phi_0$. At the detuned bias, the qubit frequency was 4.333\,GHz, and, hence, the qubit frequency detuning was $\delta \omega_{\text{q}}/2\pi \approx 3$\,MHz. In addition, Fig.\,\ref{fig3}(a) shows the values of $(\Gamma^\prime_1/2)^{-1}$ at the optimal point. The relaxation rate $\Gamma^\prime_1(\Omega_\text{R})$ was estimated from the data shown in Figs.\,\ref{fig1}(a),(b) using the equation $\Gamma^\prime_1 \approx \Gamma_1 (\omega_{\text{q}})/2+\Gamma_1 (\omega_{\text{q}}+\Omega_\text{R})/2$ obtained from Eq.\,(\ref{eq:exact}) under the assumption $\Gamma_1 (\omega_{\text{q}}-\Omega_\text{R})\approx \Gamma_1 (\omega_{\text{q}})$ (here, we neglected the contribution from defects with $\Delta_\text{TLS} < 0 $). At low Rabi frequencies, in accordance with Eq.\,(\ref{eq:eq1}), the $T_{1\rho}$ time was dominated by the term $\Gamma_{\Omega}^{-1}(\Omega_\text{R})$ due to the low-frequency $1/f$ noise of the high-power amplifier. At high Rabi frequencies, the relaxation time plateaued at the value of 100\,$\mu$s which was below the baseline level of $(\Gamma^\prime_1/2)^{-1}$. In this frequency range, the $1/f$ noise was not dominant, and the difference between background values of $T_{1\rho}$ and $(\Gamma^\prime_1/2)^{-1}$ was caused by other low-frequency noises at $\Omega_R$, and by high-frequency TLS defects with $\Delta_\text{TLS} < 0 $ and short relaxation times (broad spectra). The pronounced features in the $T_{1\rho}$ data were caused by the coupling between the qubit and high-frequency TLS defects with long relaxation times (narrow lines). As explained above, the main feature observed at the optimal point was attributed to the TLS-3 defect. At the detuned flux bias, positions of spectral features related to the defects with $\Delta_{\text{TLS}} > 0$ were shifted to lower Rabi frequencies by the value of the qubit frequency detuning $\delta \omega_{\text{q}}$. Therefore, we identified the feature observed at $\Omega_\text{R}/2\pi \approx 45$\,MHz at the detuned bias as the main one corresponding to the TLS-3 defect~[Fig.\,\ref{fig3}(a)]. As shown in Fig.\,\ref{fig3}(b), there were some discrepancies in the positions of TLS defects observed in $T_1$ and $T_{1\rho}$ measurements which could be caused by fluctuations of defect frequencies\,\cite{Mueller2015,Klimov2018,Burnett2019}. In separate spin-locking measurements, we found that positions of the spectral features were not affected by in-plane magnetic fields of up to 0.2\,mT, which indicated that the detected TLS defects were charge defects\,\cite{SM}. 

In conclusion, we demonstrated that spin-locking measurements of the driven-state relaxation of a superconducting qubit with the frequency $\omega_{\text{q}}$ can be significantly affected by the interaction with off-resonant TLS defects with the frequencies $\omega_{\text{TLS}}$ if one of the conditions $\omega_{\text{TLS}}=\omega_{\text{q}} \pm \Omega_\text{R}$ is met. Although the qubit and defects were nominally off-resonance in the laboratory frame in our experiments, the Purcell-like regime of resonant qubit-defect coupling was realized in the rotating frame by driving the qubit with the spin-locking pulse. As a result, the qubit relaxation was affected by the interaction with the defects. Thus, in addition to the effect of low-frequency noise sources reported previously\,\cite{Yan2013}, spectral features in spin-locking measurements can be caused by off-resonant high-frequency defects. Similar effects can be observed in other types of driven-state noise spectroscopy such as Rabi and rotary-echo measurements\,\cite{Bylander2011,Gustavsson2012,Yoshihara2014}. The reported qubit-defect coupling can be also interpreted in terms of Autler-Townes splitting of the qubit state as the interaction between the defect and one of the qubit dressed states\,\cite{Baur2009,Astafiev2010}. Our results demonstrate that spin-locking methods can be used to couple a superconducting qubit to other quantum systems with high-frequency transitions such as NV centers in diamond\,\cite{Saito2013}. Our work also shows that spin-locking techniques can be used to determine the spectral distribution of high-frequency defects in the vicinity of the qubit frequency, which can be especially useful for fixed-frequency qubits such as transmons.

\let\oldaddcontentsline\addcontentsline
\renewcommand{\addcontentsline}[3]{}
\begin{acknowledgments}
This work was partially supported by CREST (JPMJCR1774), JST.
\end{acknowledgments}

%
\let\addcontentsline\oldaddcontentsline
\widetext
\clearpage
\begin{center}
\textbf{\large Supplemental Materials \\ ``Driven-state relaxation of a coupled qubit-defect system in spin-locking measurements''}
\end{center}
\setcounter{equation}{0}
\setcounter{figure}{0}
\setcounter{table}{0}
\setcounter{page}{1}
\makeatletter
\renewcommand{\theequation}{S\arabic{equation}}
\renewcommand{\thefigure}{S\arabic{figure}}
\renewcommand{\bibnumfmt}[1]{[S#1]}
\renewcommand{\citenumfont}[1]{S#1}

\tableofcontents
\newpage

\section{Experimental setup}

\begin{figure}[hb]
    \centering
    \includegraphics[width=0.7\textwidth]{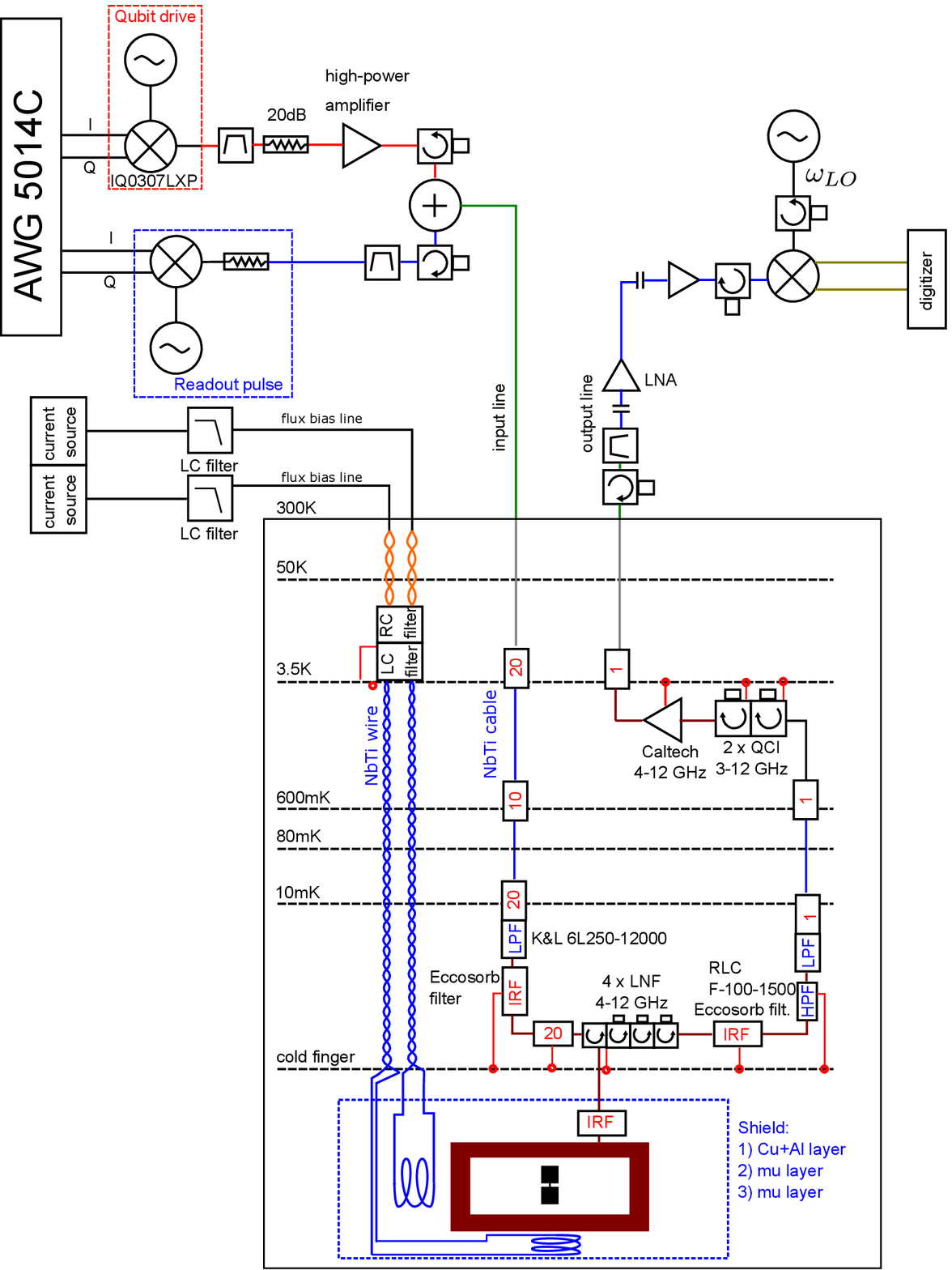}
    \caption{(color online) A schematic of the experimental setup.}
    \label{fig1S}
\end{figure}

Figure\,\ref{fig1S} shows the schematic of the experimental setup which was similar to the one used previously\,\cite{Abdurakhimov2019s}. In comparison with the previous setup, a few changes were made. First, a high-power amplifier Spacek Labs SP1018-42-27RM was inserted in the qubit drive line (the typical gain is 40 dB, with an output 1dB compression point of +30 dBm, and a noise figure of 3 dB). To protect the amplifier from the saturation, a 20 dB attenuator was placed at the input of the amplifier. Second, a custom-made low-pass LC filter (manufactured by Aivon Ltd, Finland) was added to the magnet bias line at the 3.5 K stage. Third, an additional Eccosorb filter was placed inside the magnetic shield. Fourth, an additional custom-made solenoid magnet (Suzuki Shokan Co. Ltd., Japan) was mounted inside the magnetic shield in order to apply an in-plane magnetic field.

\section{High-power amplifier noise}
\begin{figure}[hb]
    \centering
    \includegraphics{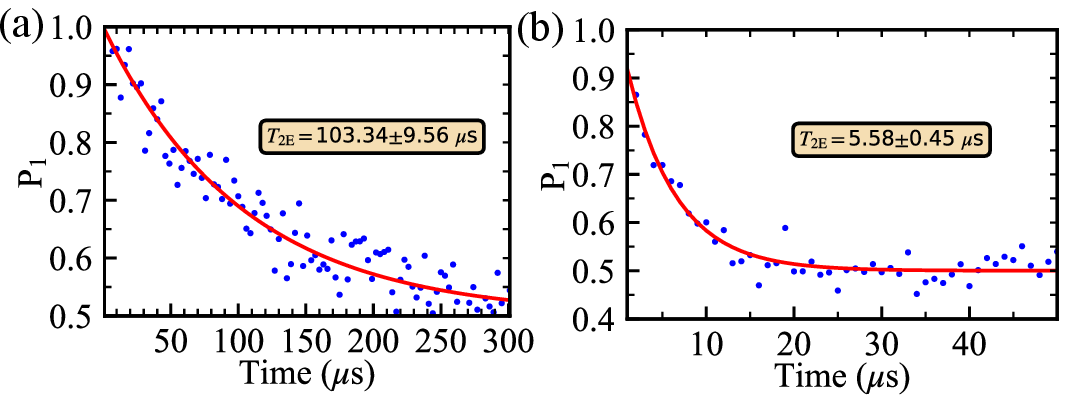}
    \caption{(color online) Results of $T_\text{2E}$ measurements: (a) without,  and (b) with the high-power amplifier connected.}
    \label{fig2S}
\end{figure}
As mentioned in the main text, qubit dephasing time $T_2$ was significantly reduced in experiments with the high-power amplifier. In measurements without the amplifier, $T_\text{2E}$ was about 100\,$\mu$s [Fig.\,\ref{fig2S}(a)]. After the installation of the amplifier, $T_\text{2E}$ reduced to approximately 6\,$\mu$s [Fig.\,\ref{fig2S}(b)].
\begin{figure}[hb]
    \centering
    \includegraphics{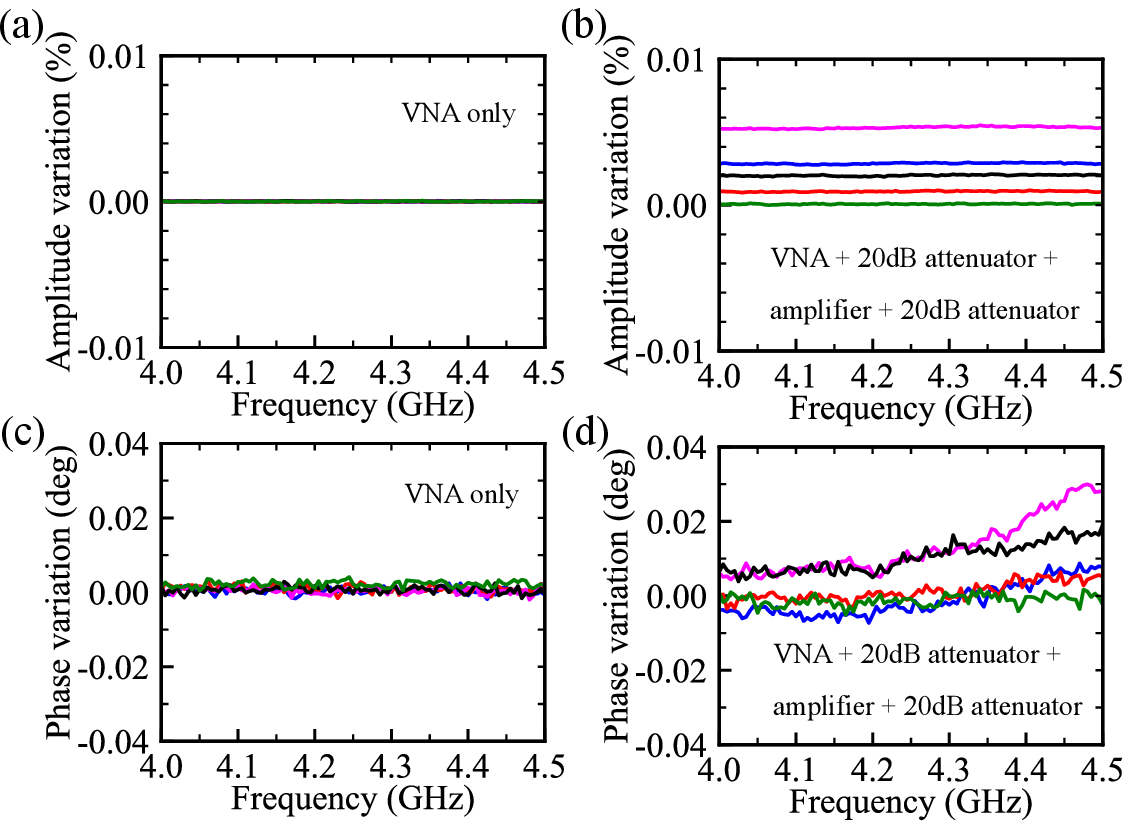}
    \caption{(color online) Measurements of amplitude and phase variations using a vector network analyzer. (a, b) Amplitude fluctuations without and with the amplifier connected, respectively. (c, d) Phase fluctuations without and with the amplifier connected, respectively. (a--d) Each plot contains 5 traces obtained in 10 successive measurements, with each trace representing the difference between the data collected in two consecutive measurements.}
    \label{fig3S}
\end{figure}
The observed deterioration of the dephasing time can be caused by the amplitude and phase noise $\delta A_N(t)$ and $\phi_N(t)$, respectively, which are introduced by the amplifier into the qubit drive signal $S=[A+\delta A_N(t)]\cos{[\omega_q t + \phi_N(t)]}$. The amplitude noise results in the qubit rotation errors. The phase noise contribution to the dephasing was discussed previously in connection with the phase noise of a local oscillator\,\cite{Ball2016}. Figure\,\ref{fig3S} demonstrates low-frequency amplitude and phase fluctuations introduced by the amplifier which were measured using a vector network analyzer (VNA). The probe power was -10 dBm, and the IF bandwidth of the VNA was 100\,Hz. To compensate for the 40 dB amplifier gain, a 20 dB attenuator was inserted at the amplifier input, and another 20 dB attenuator was mounted at the amplifier output. The amplitude (phase) variations were obtained by calculating the amplitude (phase) difference of signals recorded in two consecutive measurements. As shown in Fig.\,\ref{fig3S}, the installation of the amplifier resulted in the increase of the amplitude and phase noises. It should be noted that, in addition to random noise, a common trend was observed in phase signals in the measurements with the amplifier connected: phase variations increased with the measurement time [Fig.\,\ref{fig3S}(d)]. That increase was probably due to the heating of the amplifier.

\section{Amplitude calibration procedure}
\begin{figure}[hb]
    \centering
    \includegraphics{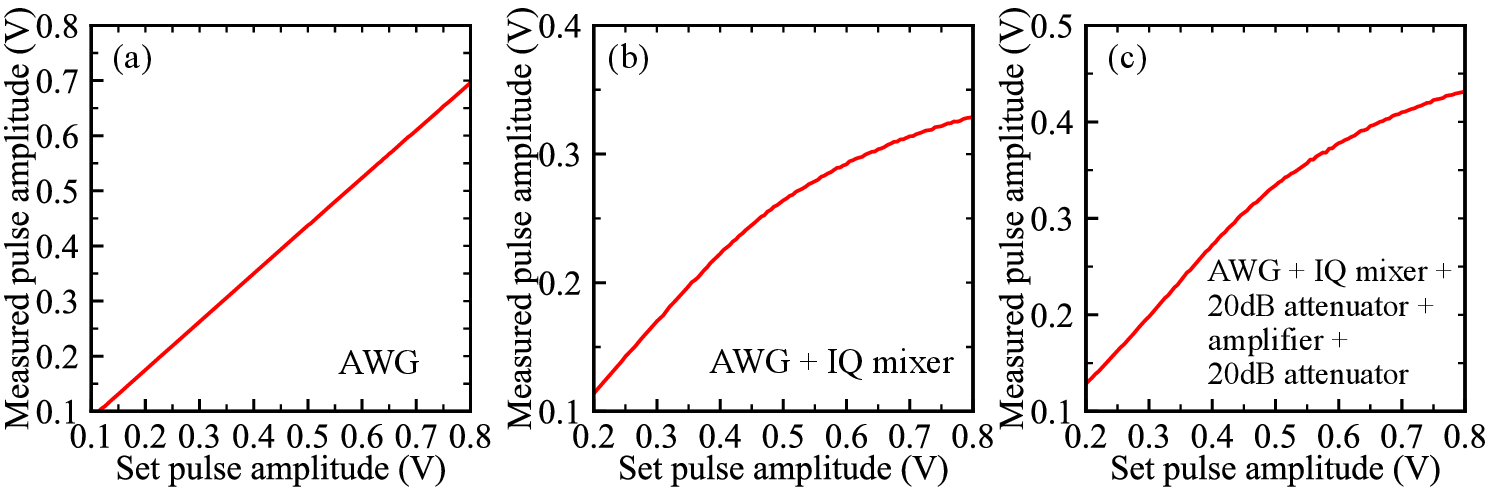}
    \caption{(color online) Drive amplitude calibration. (a) The drive pulse amplitude at the output of the AWG. (b) The pulse amplitude at the output of the IQ mixer. (c) The pulse amplitude at the output of the high-power amplifier (attenuated by a 20 dB attenuator).}
    \label{fig4S}
\end{figure}
To calibrate the qubit drive amplitude, we measured pulse amplitudes at different stages of the input drive line using a high-frequency oscilloscope Keysight DSOV204V (Fig.\,\ref{fig4S}). First, we measured the pulse amplitude at the output of the arbitrary wave generator Tektronix AWG5014C [Fig.\,\ref{fig4S}(a)]. The cable insertion loss was very small, so the output pulse amplitude was very close to the value of the pulse amplitude set in the AWG settings. Second, we measured the output pulse amplitude after the IQ mixer (Marki Microwave IQ0307LXP) [Fig.\,\ref{fig4S}(b)]. The non-linear behavior was observed due to the saturation of the mixer output power (according to the mixer datasheet, the typical input 1dB compression point is +4 dBm). Finally, we measured the pulse amplitude at the output of the high-power amplifier [Fig.\,\ref{fig4S}(c)]. To protect the input of the oscilloscope, a 20 dB attenuator was installed at the input of the amplifier, and an additional 20 dB attenuator was placed at the amplifier output. Figure\,1(d) of the main text was plotted using calibrated amplitude values extracted from the data shown in Fig.\,\ref{fig4S}(c).

\newpage
\section{Defect identification procedure}
\begin{figure}[hb]
    \centering
    \includegraphics{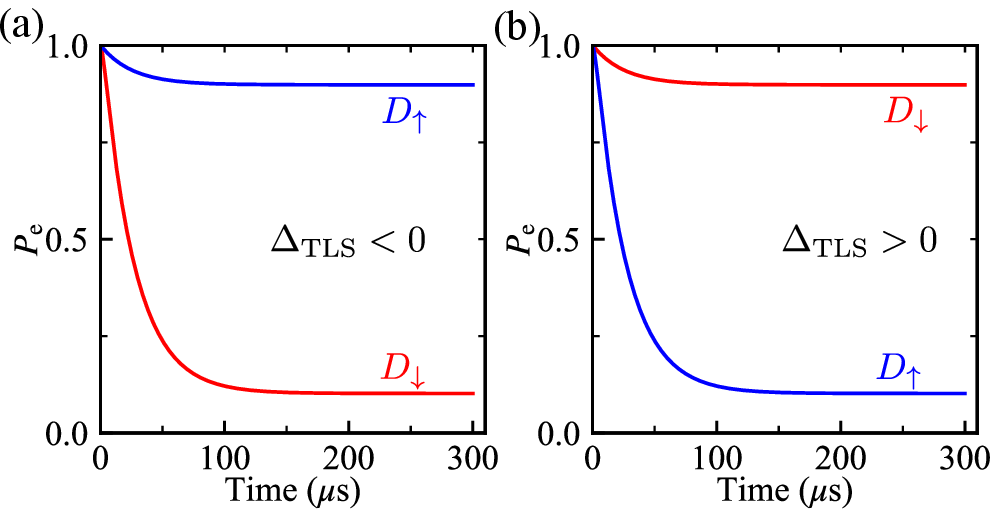}
    \caption{(color online) Numerical simulations of spin-locking experiments. (a) The driven-state relaxation of the qubit coupled to a negatively detuned defect. (b) The driven-state relaxation of the qubit coupled to a positively detuned defect.}
    \label{fig5S}
\end{figure}

\begin{figure}[hb]
    \centering
    \includegraphics{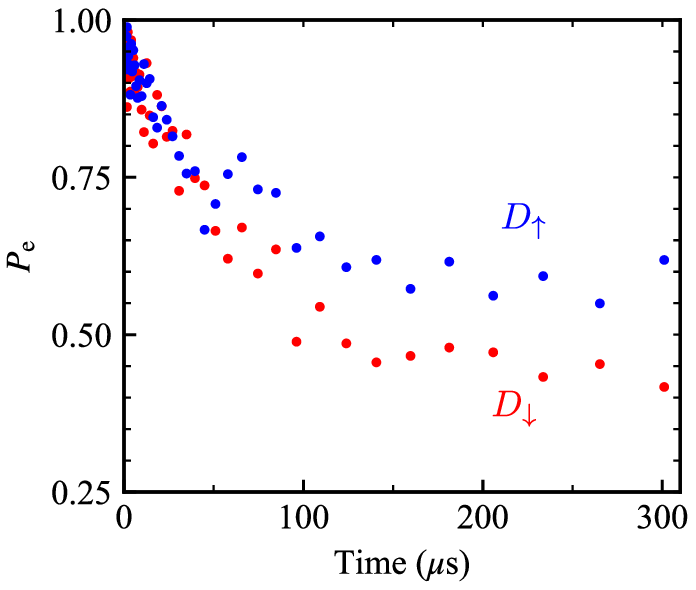}
    \caption{(color online) Results of spin-locking measurements at the optimal bias point at the Rabi frequency of 70 MHz. Since the $D_\downarrow$ response is below the $D_\uparrow$ signal, the observed signature corresponds to a negatively detuned defect.}
    \label{fig6S}
\end{figure}

Figure\,\ref{fig5S} shows results of numerical simulations of spin-locking experiments for negatively and positively detuned defects. All parameters were the same as the ones used in the simulation described in the main text, except for the sign of the TLS detuning $\Delta_\text{TLS}$. 
In each case, we plotted qubit excited-state populations $D_\uparrow$ and $D_\downarrow$ corresponding to two spin-locking sequences S1 and S2, respectively, as described in the main text. As can be easily seen from Fig.\,\ref{fig5S}, $D_\uparrow$ and $D_\downarrow$ responses for $\Delta_\text{TLS} <0$ and $\Delta_\text{TLS} > 0$ are inverted. Thus, one can say that the signatures for negatively and positively defects have different ``polarities''. For example, Fig.\,\ref{fig6S} shows the experimental raw data obtained in spin-locking measurements at the optimal point at the Rabi frequency of 70 MHz. We see that the $D_\downarrow$ signal is below the $D_\uparrow$ one, and, hence, this feature corresponds to a negatively detuned defect. An example of the raw data for a positively detuned defect is shown in Fig.\,2(d) of the main text.

\begin{figure}[hb]
    \centering
    \includegraphics{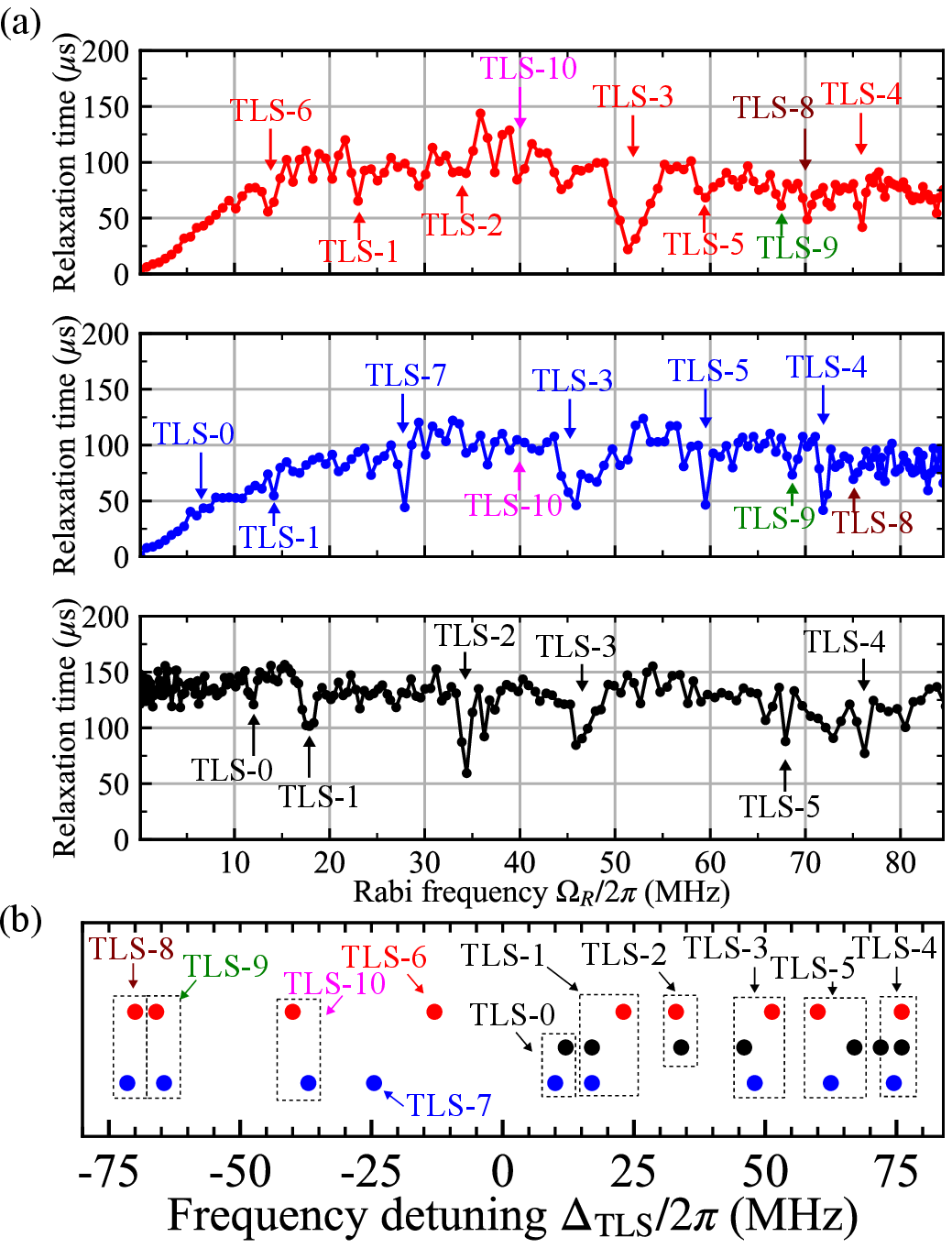}
    \caption{(color online) The extended version of Fig.\,3 of the main text. (a) The relaxation time $T_{1\rho}$ of the driven qubit state. Red dots represent $T_{1\rho}$ at the optimal bias point $\delta \Phi =0$. Blue dots correspond to $T_{1\rho}$ at the flux bias detuned from the optimal point by $\delta \Phi = 7.5\times10^{-4}\,\Phi_0$. Black dots represent the estimated values of $(\Gamma^\prime_1/2)^{-1}$ at the optimal bias point. Markers show spectral features caused by the coupling between the qubit and TLS defects in corresponding measurements. (b) Frequencies of high-frequency TLS defects determined from $T_1$ (black dots) and $T_{1\rho}$ (red and blue dots) measurements. The horizontal axis represents the frequency detuning $\Delta_{\text{TLS}} = \omega_{\text{TLS}} - \omega_{\text{q}}$. Red dots correspond to spectral features inferred from the $T_{1\rho}$ raw data obtained at the the optimal point~[Fig.2(a),(b) of the main text]. Blue dots represent defect positions determined from the $T_{1\rho}$ raw data obtained at the detuned bias [Fig.\,\ref{fig8S}(a),(b)], with the corresponding values $\Delta_{\text{TLS}}$ being increased by the value $\delta \omega_{\text{q}}$. Dashed-line rectangles indicate spectral features caused by the same TLS defect.}
    \label{fig7S}
\end{figure}

Using the described approach, we identified multiple negatively and positively detuned defect signatures in our measurements [Fig.\,\ref{fig7S}]. It should be noted that it was easier to distinguish defect signatures from the noise using raw data [e.g., Fig.\,\ref{fig6S}, and Fig.\,\ref{fig8S}(a),(b)] than using phase-cycled data [e.g., Fig.\,\ref{fig8S}(c), and top two panels in Fig.\,\ref{fig7S}(a)].

\clearpage
\section{Measurements at the detuned bias}
\begin{figure}[hb]
    \centering
    \includegraphics{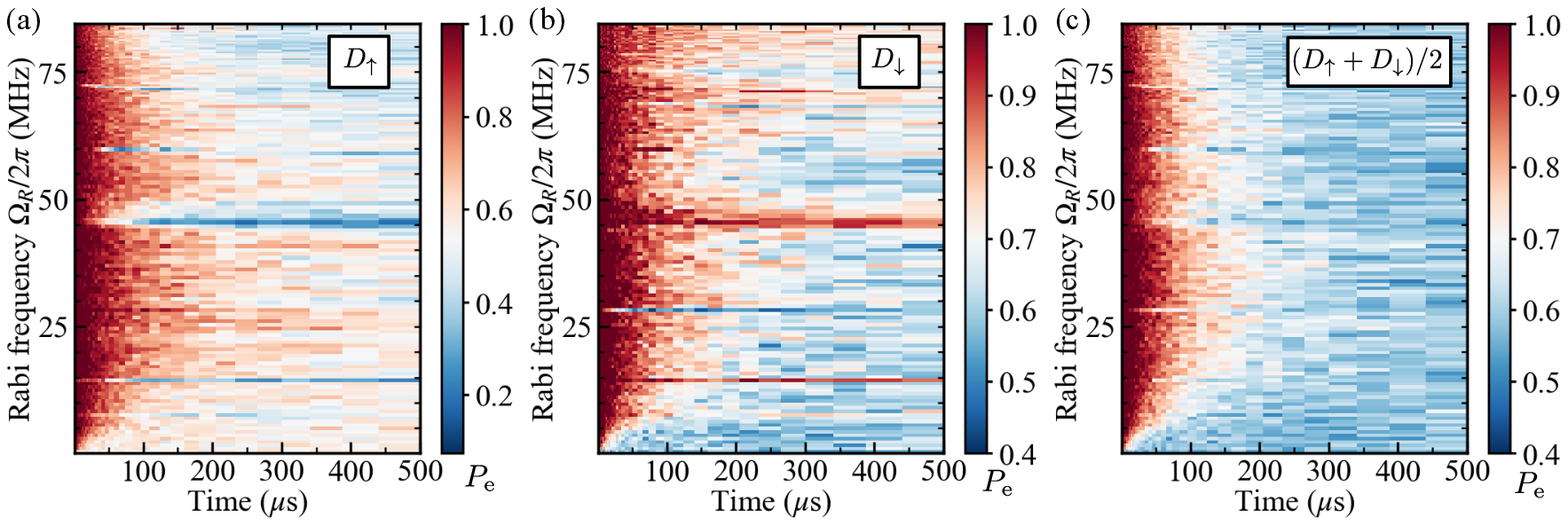}
    \caption{(color online) Results of spin-locking spectroscopy at the flux bias detuned from the optimal point by $\delta \Phi = 7.5\times10^{-4}\,\Phi_0$ (the qubit frequency detuning was $\delta \omega_{\text{q}}/2\pi \approx 3$\,MHz). (a) The data set $D_\uparrow$ represents results obtained using the pulse sequence $(-\frac{\pi}{2})_Y - \text{SL}_X - (-\frac{\pi}{2})_Y$. (b) The data set $D_\downarrow$ corresponds to results obtained using the pulse sequence $(+\frac{\pi}{2})_Y - \text{SL}_X - (+\frac{\pi}{2})_Y$. (c) The arithmetic mean $P=(D_\uparrow+D_\downarrow)/2$.}
    \label{fig8S}
\end{figure}

In Fig.\,\ref{fig8S}, we show the raw experimental data obtained in spin-locking measurements at the detuned bias, which was used to plot Fig.\,3 of the main text.

\section{Experiments with the applied in-plane magnetic field}
\begin{figure}[hb]
    \centering
    \includegraphics{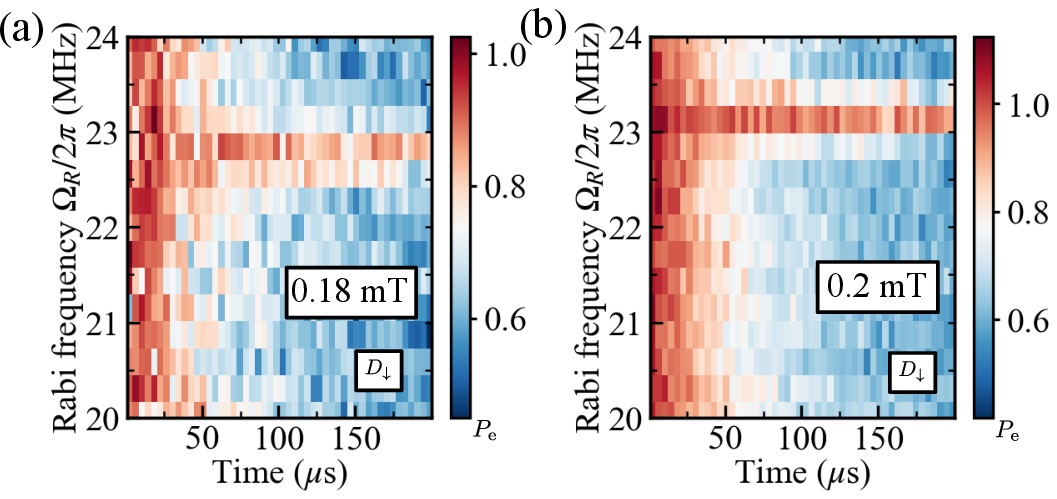}
    \caption{(color online) Results of spin-locking spectroscopy at the optimal point with the in-plane magnetic field using the pulse sequence $(+\frac{\pi}{2})_Y - \text{SL}_X - (+\frac{\pi}{2})_Y$. (a) Measurements with the in-plane magnetic field of 0.18 mT. (b) Measurements with the in-plane magnetic field of 0.2 mT.}
    \label{fig9S}
\end{figure}
In Ref.\,\cite{Yan2013s}, it was suggested that features in spin-locking measurements can be caused by low-frequency electron spin defects with the g-factor $g\approx2$ and the gyromagnetic ratio $\gamma/2\pi \approx 28$\,MHz/mT. In separate experiments, we performed spin-locking measurements in applied in-plane magnetic fields of 0.18 mT and 0.2 mT (Fig.\,\ref{fig9S}). The perpendicular bias field used to tune the qubit to the optimal point was about 0.02 mT which was much smaller than the in-plane field. The qubit frequency was about 4.395\,GHz, and the main feature observed in spin-locking measurements was located at the Rabi frequency of about 23 MHz (we should note that both qubit and TLS frequencies fluctuated in different experimental runs due to thermal cycling). For an electron spin defect with $\gamma/2\pi \approx 28$\,MHz/mT, we would expect to see the features at $\Omega_\text{R}/2\pi\approx 5$\,MHz at the applied magnetic field of 0.18 mT and $\Omega_\text{R}/2\pi\approx5.6$\,MHz at the field of 0.2 mT. Even if we assume that the magnet constant was different from the estimated value of 2\,$\mu$T/mA, a 10\% increase of the defect frequency should be observed in the measurements made with magnet currents of 90\,mA and 100\,mA. However, in the experiment, we found that the defect signature only slightly fluctuated near the Rabi frequency of 23 MHz. Therefore, we can conclude that the defect signature observed in our measurements cannot be explained by Zeeman splitting of a spin defect.


\begin{thebibliography}{40}
\bibitem [{thi()}]{third_note}%
  \BibitemOpen
  \href@noop {} {}\bibinfo {note} {At a given Rabi frequency, the steady levels
  of the expectation value $\langle\sigma_x^\text{(1)}\rangle$ in the end of
  the spin-locking pulse $\text{SL}_X$ were the same for S1 and S2
  measurements.}\BibitemShut {Stop}%
\bibitem [{sec()}]{second_note}%
  \BibitemOpen
  \href@noop {} {}\bibinfo {note} {For defects with $\Delta_\text{TLS}<0$, the
  ground state in the laboratory frame corresponds to the excited state in the
  rotating frame. Therefore, such defects can be considered as ``hot''
  defects.}\BibitemShut {Stop}%
\end{thebibliography}
%

\end{document}